\documentclass[final,5p,times]{elsarticle}

\usepackage{amssymb,amsmath,amsfonts}
\usepackage{graphicx}
\usepackage{subcaption}
\usepackage[normalem]{ulem} 
\usepackage{comment}
\usepackage{xcolor}
\usepackage{enumitem}
\usepackage{fancyvrb}

\usepackage{url}
\usepackage{hyperref}
\usepackage{doi}
\newcommand{\dd}{\mathrm d}

\AtBeginDocument{%
}

\biboptions{square,comma,sort&compress}

\newcommand{\jls}[1]{\textcolor{magenta}{\ifmmode\text{\cancel{\ensuremath{#1}}}\else\sout{#1}\fi}}

\journal{Computer Physics Communications}

\begin{document}

\begin{frontmatter}



\title{DYNAMITE: A high-performance framework for solving\\ Dynamical Mean-Field Equations} 


\author[1]{Johannes Lang}

\author[2,3]{Vincenzo Citro}

\author[3,5]{Luca Leuzzi}

\author[5,3,6]{Federico Ricci-Tersenghi}

\affiliation[1]{
organization={Institut f\"ur Theoretische Physik, Universit\"at zu K\"oln}, 
addessline={Z\"ulpicher Stra{\ss}e 77},
city={Cologne},
postcode={50937},
country={Germany}}

\affiliation[2]{
organization={DIIN, Universit\'a di Salerno}, 
addressline={Via Giovanni Paolo II 132},
city={Fisciano},
postcode={84084},
country={Italy}}

\affiliation[3]{
organization={CNR--Nanotec, unit\`a di Roma}, 
addressline={P.le Aldo Moro 5},
city={Rome}, 
postcode={00185},
country={Italy}}

\affiliation[5]{
organization={Dipartimento di Fisica, Sapienza Universit\'a di Roma},
addressline={P.le Aldo Moro 5},
city={Rome}, 
postcode={00185},
country={Italy}}

\affiliation[6]{
organization={INFN, Sezione di Roma I},
addressline={P.le Aldo Moro 5},
city={Rome}, 
postcode={00185},
country={Italy}}


\begin{abstract}
Understanding the dynamics of systems evolving in a complex and rugged energy landscape is fundamental in various fields, including physics, economics, biology, and computer science. Disordered mean-field models are a powerful framework for studying these processes, as exact Dynamical Mean-Field Equations (DMFE) can be derived.
However, the solution to the DFME---a set of coupled integral-differential equations in two-time functions---has always been a truly challenging numerical task.

Until now, the solution to the DMFE in the large times limit has been obtained in two ways: analytically, through the Cugliandolo--Kurchan ansatz that relies on some hypothesis (e.g., the weak ergodicity breaking hypothesis, which has been numerically shown not to hold in certain cases~\cite{Bernaschi2020, Folena2020, Folena2023, citro2025strong, Lang2025}); via numerical integrations that reach times $O(10^3)$ reliably and longer times only using approximations which are not well under control.
In practice, currently, there is no method to solve generic DMFE for very long times, which imposes significant limitations on our understanding of slow dynamical processes occurring on complex energy landscapes.

In this work, we present \textsc{Dynamite} (DYNAmical Mean-fIeld Time Evolution solver), a high-performance computational framework to solve DFME that explores unprecedented time scales, $t=O(10^7)$, through a combination of two-dimensional non-uniform interpolation, adaptive time steps, and numerical `renormalization' of the system memory. The algorithm constructs compact interpolation stencils over the causal domain, enabling the evaluation of history integrals with high-order accuracy. Asymptotic running times are linear, and the memory required is sublinear in the total integration time. Stability and precision are maintained via an adaptive Runge--Kutta integrator and a periodic sparsification procedure that coarsens the past while preserving monitored observables within prescribed tolerances. 

\textsc{Dynamite} achieves orders-of-magnitude speedups compared to uniform-grid approaches while retaining quantitative accuracy and reproducibility across CPU and GPU architectures. Representative benchmarks on glassy mean-field models, including the mixed spherical $p$-spin system, demonstrate reliable access to aging and relaxation regimes previously inaccessible to direct simulation. The framework provides a reproducible, extensible, and open foundation for the numerical study of long-memory dynamical systems across statistical physics and beyond.
\vspace{5mm}

Program summary.

Program title: DYNAMITE

Licensing provisions: Apache License 2.0

Programming language: C++ (CUDA-enabled build; CPU-only fallback available)

Operating system: Linux, macOS (CPU mode), HPC systems with NVIDIA GPUs 

Nature of problem: Long-time non-equilibrium dynamics in glassy systems governed by dynamical mean-field equations are computationally demanding with standard time-stepping approaches.

Solution method: Numerical renormalization of two-time dynamics using adaptive interpolation grids for correlation and response functions combined with sparsification and GPU-accelerated kernels.

External routines/libraries: CUDA toolkit (optional), OpenMP, HDF5 (optional), CMake.

Program documentation: \href{https://dmft-evolution.github.io/DYNAMITE/}{\texttt{https://dmft-evolution.github.io/DYNAMITE/}}

\end{abstract}




\end{frontmatter}


\section{Introduction}
The aging dynamics is a characteristic feature of the out-of-equilibrium evolution of frustrated complex systems, including spin glasses, structural glasses, neural networks, inference problems, and quantum many-body systems.
In the dynamical regime called aging, the system evolution becomes increasingly slow as time progresses, which requires integrating the dynamics over longer and longer timescales.
Despite decades of intense research, a general analytical description of the aging dynamics remains elusive, and the numerical methods to integrate the dynamics at long times play a crucial role.

From a theoretical standpoint, mean-field spin glass models are considered the simplest prototypes for studying the aging dynamics. In particular, the mean-field spin glass models with a Random First Order Transition (RFOT) show the typical phenomenology of glass formers \cite{Kirkpatrick1987, Kirkpatrick1989, franz2001ferromagnet, Lubchenko2007, Leuzzi07b, ricci2010being}.

Among these spin glass models, the so-called spherical $p$-spin model \cite{Crisanti1992,Crisanti1993,Ferrari2012} stands out as a simple, but rich paradigmatic example.
Thanks to the fully-connected topology of the interactions among the $N$ variables (which are real numbers subject solely to a global spherical constraint, but this can even be relaxed), the stochastic dynamics of the system, in contact with a thermal bath a temperature $T$, can be exactly integrated in the large $N$ limit, providing a set of integro-differential equations in two-times correlation and response functions, $C(t,t')$ and $R(t,t')$, called Dynamical Mean-Field Equations (DMFE).

In their seminal work, Cugliandolo and Kurchan \cite{cugliandolo1993analytical} introduced an ansatz for solving the DMFE analytically in the limit of very large times. At the core of their approach lies the concept of \emph{weak ergodicity breaking} (WEB): during aging, the system continues to evolve ever more slowly while exploring an unbounded manifold of marginal states, progressively forgetting any configuration reached at finite times. This separation between short- and long-time dynamics enables a closed analytical description and has long been considered a cornerstone of mean-field glassy dynamics.

However, accumulating numerical and analytical evidence suggests that this picture is fragile and does not extend beyond the \emph{pure} spherical $p$-spin model (see definition below). Already in the simplest generalization, the \emph{mixed} spherical spin models, whose Hamiltonian includes interactions involving different numbers of spins and whose statics and equilibrium dynamics are known \cite{Nieuwenhuizen1995, Crisanti2006, Crisanti2007, Crisanti07b, Sun_2012}, the WEB hypothesis appears to fail. 
Indeed, the numerical integration of the DMFE for mixed spherical spin models clearly indicates the emergence of \emph{strong ergodicity breaking} (SEB), whereby the system never completely forgets its past and the aging dynamics remain confined within a shrinking region of configuration space \cite{Folena2020, Folena2023, Lang2025, citro2025strong}.
The same SEB effects have been observed in other systems \cite{Bernaschi2020} and are believed to be a general property of mean-field aging systems.

There is, then, a sharp gap between equilibrium knowledge and out-of-equilibrium dynamical understanding. Bridging this gap requires access to the true full-time regime of the DFME. 
The main obstacle is technical but fundamental. Solving the DMFE entails evolving two-time correlation and response functions, leading to a memory cost that scales quadratically and a computational cost that scales cubically with the maximum integration time. As a consequence, early numerical studies of mixed spherical spin models based on uniform discretization grids were limited to times of order $10^3$, insufficient to reliably probe the long-time aging regime \cite{Folena2020}. Although these studies already revealed striking qualitative differences between pure and mixed models, they could not decisively test the assumptions underlying the Cugliandolo--Kurchan ansatz.

The primary strategy adopted in the past to overcome these scaling limitations was the use of a discretization time step that grows during the aging dynamics \cite{Kim2001, Andreanov09, Mannelli2020}. However, this approach has proven unstable for mixed spherical models \cite{folena2020thesis}, where uncontrolled error accumulation prevents reliable integration beyond the initial transient. Accessing the deep aging regime, therefore, requires a more sophisticated non-uniform grid structure capable of maintaining accuracy across the full history. Recent work has demonstrated that such approaches can reach times of order $10^7$~\cite{Lang2025,citro2025strong}; in the present work, we describe in detail the integration scheme and algorithmic principles underlying the method introduced in Ref.~\cite{Lang2025}.

Accessing long timescales is essential to capture physical phenomena that only emerge deep in the aging regime. One practical example may be very illuminating. Experiments on spin glass materials (that are intrinsically out of equilibrium and display aging on any experimentally accessible time scale at low enough temperatures) have revealed surprising phenomena, called rejuvenation and memory \cite{jonason1998memory}.
For decades, attempts to reproduce such phenomena in numerical simulations were unsuccessful \cite{picco2001chaotic, berthier2002geometrical, takayama2002numerical, maiorano2005edwards, jimenez2005rejuvenation}, the main reason being that the probed timescales were too short.
Only recently, thanks to the use of the special-purpose Janus II computer \cite{baity2014janus} that makes very long timescales accessible, both rejuvenation and memory phenomena were observed in the out-of-equilibrium aging dynamics of spin glass models \cite{baity2023memory}.
In mean-field fully-connected models, Monte Carlo simulations are highly inefficient; thus, a numerical tool to solve DMFE at long times is strictly needed.

Beyond the study of the aging regime in spin glass models, the solution in the long-time limit to DMFE (or equations very similar in their form) is relevant in many other fields. Two-time dynamical field-theory equations analogous to those of mixed spherical spin models arise naturally in non-equilibrium quantum systems described within the Keldysh formalism \cite{Anous2021, Sieberer2016, Keldysh1964} and in classical stochastic dynamics formulated through the Martin–Siggia–Rose approach and its subsequent developments \cite{Martin1973, Janssen1976a, DeDominicis1976, DeDominicis1978} (that lead to the DMFE for spin glasses).
Closely related dynamical equations also appear in the solutions to models relevant to the understanding of neural networks \cite{Crisanti1987, Crisanti1988, Sompolinsky1988, Crisanti2018, Fournier2023, Badalotti2025, Fournier2025, montanari2025dynamical}, inference problems formulated as high-dimensional stochastic processes \cite{mannelli2019passed, mignacco2020dynamical}, and Hopfield-type neural networks evolving under Glauber dynamics \cite{Rieger1988, Bolle2003}.

Motivated by these challenges, this work illustrates in detail a new numerical integration scheme for the two-time DMFE, which leverages the inherent multi-scale nature of aging dynamics \cite{Lang2025}. By systematically discarding irrelevant past information via adaptive time rescaling, with a fixed computational cost per time step, the method dramatically reduces memory requirements while preserving stability and accuracy at very long times. 
The code is available on a dedicated webpage \cite{DYNAMITE2026}, where complete documentation is provided.
It is provided in a form suitable to solve the out-of-equilibrium dynamics in spherical mixed spin models, but it can be easily adapted to any of the problems listed above.
Given that the new integration scheme can reach, especially when run on GPUs, time scales that are several orders of magnitude larger than those accessible with previous methods, we expect \textsc{Dynamite} to have an impact in several areas of research.

\section{Theoretical framework}
\label{sec:thframework}

We consider interacting degrees of freedom evolving under stochastic or quantum dynamics. In systems with fully-connected interactions, or more generally in large-$N$ mean-field limits, the dynamics becomes self-averaging in the thermodynamic limit ($N\to\infty$). As a consequence, the evolution can be formulated as coupled integro-differential equations for the two-time correlation and response functions, $C(t,t')$ and $R(t,t')$. The generic form of these Dynamical Mean-Field Equations (DMFE) is the following
\begin{eqnarray}
    \partial_t C(t,t') &\!\!\!=& \!\!\!\!\int_0^t \dd s\, K[C,R](t,s)\,C(s,t') + \nonumber\\
    && \!\!\!\!\int_0^{t'} \dd s\, D[C,R](t,s)\,R(t',s) + F_C(t,t')\;,\label{eq:genericDMFE} \\
	\partial_t R(t,t') &\!\!\!=& \!\!\!\delta(t-t') + \int_{t'}^t \dd s\, K[C,R](t,s)\,R(s,t')\;,\nonumber
\end{eqnarray}
where $F_C(t,t')$ represents explicit driving or noise originating from the environment, while the memory kernels $K(t,t')$ and $D(t,t')$ encode the self-consistent feedback generated by interactions within the system. In general, these kernels are functionals of the dynamical state through $C$ and $R$, reflecting the fact that the collective behavior of the system self-consistently generates the effective dynamics.

Equations of this type arise in a broad range of contexts. 
They appear in dynamical mean-field descriptions of classical spin glasses and structural glasses, as well as in quantum impurity problems and in the dynamical mean-field theory of strongly correlated electrons~\cite{Georges1996}. 
Closely related structures also emerge in large-$N$ quantum field theories and other non-equilibrium many-body systems~\cite{Aoki2014}. In these settings, the DMFE can be derived from microscopic stochastic dynamics (e.g., Langevin equations in contact with a thermal bath) or from quantum non-equilibrium formulations in which a local degree of freedom is coupled self-consistently to an effective bath.

A direct numerical treatment of Eqs.~\eqref{eq:genericDMFE} on an equidistant time grid with step size $\delta t$ up to a maximal time $T_{\text{sim}}=N\delta t$ requires $N$ time steps. Since the evaluation of the memory integrals at each step involves sums over the full past history, the overall computational cost typically scales as $\mathcal{O}(N^3)$, with a corresponding $\mathcal{O}(N^2)$ memory footprint. This rapidly becomes prohibitive in situations where
\begin{enumerate}[label=(\roman*)]
\item the dynamics is slow, necessitating access to very long times together with the retention of the full two-time history, and/or
\item the evaluation of the kernels $K$ and $D$ is itself computationally expensive.
\end{enumerate}

The main aim of the present work is to mitigate this unfavorable scaling by combining high-order interpolation schemes with irregular, partially adaptive time grids and precomputed interpolation weights, thereby substantially reducing the number of required kernel evaluations while maintaining controlled accuracy.

While we focus on the spherical mixed $p$-spin model to illustrate our approach, the underlying strategy—high-order interpolation on non-uniform grids—applies to any DMFE of the form \eqref{eq:genericDMFE}, including quantum dynamics, which can be handled by augmenting the real-time evolution with a finite imaginary-time strip that encodes the thermal initial state \cite{Kadanoff_book, Aoki2014}; similarly, higher-order derivatives can be handled within the same framework by a standard reformulation as a first-order system.

\subsection{The spherical mixed $p$-spin model}
\label{sec:p-spin}

The spherical mixed $p$-spin model is defined by a random Hamiltonian~\cite{Nieuwenhuizen1995}
\begin{equation}\label{Hmpsin}
{\cal H} = - \sum_{p=2}^{\infty} \sum_{j_1<\cdots<j_p} \alpha_p \,J_{j_1\cdots j_p} \prod_{k=1}^p s_{j_k}\,,
\end{equation}
where the spins $s_r$, $r=1,\dots,N$, are real variables subject to the spherical constraint $\sum_{r=1}^N s_r^2=N$. The couplings $J_{j_1\cdots j_p}$ are independent Gaussian random variables with variance $\overline{J_{j_1\cdots j_p}^2}=p!/(2N^{p-1})$, and the coefficients $\alpha_p$ control the relative weight of the different $p$-body interactions.

Equivalently, the Hamiltonian can be viewed as a centered Gaussian random
function on the sphere. It is fully characterized by its covariance function
\begin{equation}
\overline{{\cal H}[\sigma]{\cal H}[\tau]}=N f(q)\;,
\end{equation}
with the covariance function
\begin{equation}
f(q)=\sum_p \frac{\alpha_p^2}{2} q^p \;, \qquad
q=\frac{1}{N}\sum_i \sigma_i\tau_i \, .
\end{equation}
In the present work, we restrict ourselves to mixtures involving two
interaction orders,
\begin{equation}
f(q)=\lambda q^p+(1-\lambda)q^{s}\;,
\end{equation}
which correspond to the most commonly studied mixed spherical $p$-spin models in the literature.

The relaxational dynamics in contact with a thermal bath at temperature $T$ is generated by the Langevin equations
\begin{equation}
\partial_t s_j(t)=-\mu(t)s_j(t)-\frac{\partial {\cal H}}{\partial s_j}+\xi_j(t)\;,
\end{equation}
where $\xi_j$ is a Gaussian thermal noise with variance
\begin{equation}
    \langle\xi_j(t)\xi_k(t')\rangle=2T\,\delta_{jk}\,\delta(t-t')\;,
\end{equation}
being $\langle\cdot\rangle$ the thermal average. $\mu(t)$ enforces the spherical constraint.

In the thermodynamic limit, the disorder-averaged dynamics is fully described by the two-time correlation and response functions,
\begin{align}
C(t,t')&=\frac{1}{N}\sum_i \overline{\langle s_i(t)s_i(t')\rangle}\;,\\
R(t,t')&=\frac{1}{N}\sum_i \frac{\partial \overline{\langle s_i(t)\rangle}}{\partial h_i(t')}\;,
\end{align}
where $h_i$ denotes an external field coupled to $s_i$, and the overline represents the average over the random couplings.
By causality, $R(t,t')$ is non-zero only if $t>t'$.
Using standard functional techniques~\cite{Martin1973,DeDominicis1978,kirkpatrick1987p}, one finds that $C$ and $R$ obey closed DFME of the generic form~\eqref{eq:genericDMFE}, with kernels determined by covariance function $f$.

Explicitly, the kernels entering Eq.~\eqref{eq:genericDMFE} are given by
\begin{align}
K(t,s)&=f''(C(t,s))\,R(t,s)\;,\\
D(t,s)&=f'(C(t,s))\;,
\end{align}
where primes denote derivatives of $f$, \textit{i.e.} $f'(x)=\mathrm{d}f/\mathrm{d}x$.
The forcing term reads
\begin{equation}
F_C(t,t')=\beta_0 f'(C(t,0))\,C(t',0)\;,
\end{equation}
where $\beta_0$ is fixed by the initial condition, which is assumed to be chosen according to the Gibbs distribution at temperature $1/\beta_0$: e.g, $\beta_0=0$ for a random initial condition.
The Lagrange multiplier $\mu(t)$ is fixed self-consistently by the spherical constraint $C(t,t)=1$,
\begin{align}
\mu(t)=T+\int_0^t \dd s\,\Big[K(t,s)C(t,s)+D(t,s)R(t,s)\Big] +F_C(t,t)\;.
\end{align}

After solving the system of coupled DMFE, the energy density is given by
\begin{equation}
E(t)=-\int_0^t \dd s\, f'(C(t,s))R(t,s)-\beta_0 f(C(t,0)).
\end{equation}

In the following sections, we use the spherical $p$-spin model as a demanding benchmark for our algorithm solving two-time DFME with long memory. On the one hand, the exact analytical solution to the \emph{pure} $p$-spin model \cite{cugliandolo1993analytical}, i.e., when $f(q)=q^p$, will be used to test the algorithm accuracy; on the other hand, the numerical solution to the \emph{mixed} $p$-spin model, i.e., when $f(q)$ has more than one monomial, reveals a new physical off-equilibrium behavior which was unknown before.

\subsection{Analytical structure and properties}\label{sec:EOM_properties}

Equations of the form~\eqref{eq:genericDMFE} possess several structural features that directly dictate the requirements for a robust numerical solver. First, the evolution is strictly causal, with memory integrals involving only past times $s \le t$. Consequently, the natural computational domain for the two-time fields is the causal triangle
\[
\mathcal{T} = \{(t,t')\,:\,0\le t'\le t\le T_{\mathrm{sim}}\},
\] 
where the simulation time $T_{\mathrm{sim}}$ may be many orders of magnitude larger than the microscopic scale $\tau_{\mathrm{micro}}$. Resolving the dynamics over these vast intervals while maintaining the full two-time history is the primary challenge addressed by our algorithm.

A defining characteristic of many-body dynamics is the presence of widely separated time scales. While microscopic degrees of freedom evolve on the scale of $\tau_{\mathrm{micro}}$, the emergence of collective modes, prethermalized states, or glassy aging can lead to relaxation occurring over macroscopic scales. This multiscale structure renders uniform time-stepping prohibitively expensive for long-time simulations. Instead, it necessitates the use of irregular or adaptive grids that provide high resolution near the diagonal to capture fast transients, while employing a coarser sampling at large time separations where the evolution is significantly slower.

Furthermore, the two-time fields often exhibit singular or constrained behavior near the diagonal $t=t'$. The response function $R(t,t')$ typically contains a distributional contribution---such as a $\delta$-function or a finite jump---reflecting the instantaneous response of the system. Additionally, many physical models impose local constraints, such as the spherical constraint $C(t,t)=1$ or specific normalization sum rules common in quantum systems. These features require discretization schemes that can accurately incorporate such boundary conditions without introducing numerical instabilities.

The self-consistent feedback inherent in dynamical mean-field descriptions is encoded in the memory kernels $K$ and $D$. These are generally nonlinear functionals of the dynamical state $(C, R)$, meaning the kernels at time $t$ depend on the values of the fields at all previous times. Evaluating these integrals requires accurate interpolation of past values at specific quadrature nodes. The precision and efficiency of this interpolation are critical: to reach very late times, the computational overhead of both the integration step and the history retrieval must be carefully controlled to avoid the cubic scaling typical of naive solvers.

Collectively, causality, state-dependent kernels, and multiscale relaxation define the numerical landscape of Eqs.~\eqref{eq:genericDMFE}. The central difficulty lies in resolving high-frequency local dynamics near the diagonal while simultaneously retaining and processing information over long time separations, a task that becomes especially demanding when $T_{\mathrm{sim}} \gg \tau_{\mathrm{micro}}$. In the next section, we quantify the computational cost of straightforward discretizations and outline the resulting challenges for practical long-time simulations.

\subsection{Numerical strategies for two-time DFME}

A direct discretization of the causal triangle on a uniform $N\times N$ grid provides a natural baseline for solving Eqs.~\eqref{eq:genericDMFE}, but it is poorly matched to their multiscale structure. Uniform meshes allocate equal resolution to all time separations, despite the fact that the fields typically vary rapidly only near the diagonal and evolve much more slowly at large time differences. Consequently, straightforward implementations incur $\mathcal{O}(N^3)$ computational cost and $\mathcal{O}(N^2)$ memory usage, which quickly becomes prohibitive for long-time simulations.

Within the class of deterministic solvers, Kim and Latz~\cite{Kim2001} introduced an early acceleration strategy based on a dyadic compression of the stored history. While this approach can significantly extend accessible timescales for the pure $p$-spin model, it lacks the stability and accuracy required for more complex dynamical landscapes \cite{folena2020thesis}. This limitation is rooted in a general feature of Eqs.~\eqref{eq:genericDMFE}: the gradients of the correlation and response functions do not decay uniformly in the $(t,t')$ plane. Even in regimes where the global relaxation appears slow, there exist regions where the fields vary rapidly. Consequently, the limited flexibility in the placement of the grid points of dyadic schemes leads to significant integration errors that accumulate over the course of the evolution, eventually destabilizing the solution.

A distinct alternative avoids the direct integration of the DFME by simulating the equivalent Langevin (or Metropolis) dynamics through stochastic sampling \cite{Bernaschi2020}. Its main drawback is that high-precision results require averaging over a large number of disorder realizations. This becomes increasingly expensive when one seeks to resolve subtle critical properties or reach the extreme timescales necessary to capture deep-aging phenomena.

The strategy developed here, and previously employed in Ref.~\cite{Lang2025}, overcomes these hurdles by representing the two-time fields on a non-equidistant causal grid. This allows the solver to maintain high local resolution wherever the evolution is rapid---effectively addressing the non-uniformity of the gradients---while employing coarser sampling where the dynamics are slow. By combining this grid structure with high-order interpolation and adaptive time integration, we achieve a scaling that allows for simulations up to $T \sim 10^7$ with controlled precision.

\section{Numerical methods and algorithmic implementation}
\label{sec:algorithm}

This section describes the numerical scheme underlying \textsc{Dynamite} and its concrete realization. Starting from the causal two-time formulation introduced in Sec.~\ref{sec:thframework}, the solver combines a non-uniform discretization of the causal triangle with high-order interpolation, explicit adaptive Runge--Kutta integration, and controlled thinning of the stored history. Together, these ingredients enable stable long-time integration of the dynamical mean-field equations while keeping computational and memory costs manageable.

We first summarize the logical structure of a single time-advance step, before discussing each component in detail.

\subsection{Algorithmic overview}

The solver propagates the two-time fields $\mathcal{C}(t,\theta)\equiv C(t,t'=\theta t)$ and $\mathcal{R}(t,\theta)\equiv R(t,t'=\theta t)$ forward in $t$, while preserving causality and controlling approximation errors arising from interpolation, quadrature, and time integration. Conceptually, each accepted time step consists of the following operations:

\begin{enumerate}
\item Propose a time increment $\Delta t$ from the adaptive ODE controller.
\item Using the fixed $\theta$ grid and the stored history in $t$, interpolate the required values of $C$ and $R$ at the quadrature nodes entering the memory integrals.
\item Assemble the memory convolutions from these interpolated values and evaluate the resulting right-hand sides of the dynamical equations.
\item Advance $C$ and $R$ with an explicit Runge--Kutta method and compute the associated error estimate.
\item Accept or reject the proposed step based on the Runge--Kutta error; upon acceptance, append the new time slice to the history.
\item Periodically apply controlled thinning of the stored past and write checkpoint data.
\end{enumerate}

All operations acting on the relative-time coordinate are performed on the fixed $\theta$ grid, while the evolution in $t$ is handled by the adaptive integrator. The separation between these two roles allows interpolation weights and quadrature coefficients associated with $\theta$ to be precomputed once, so that each time step reduces to repeated local interpolations and weighted sums over the stored history.

The following subsections describe in detail the construction of the two-dimensional grid, the interpolation strategy, the evaluation of the convolution integrals, the adaptive Runge--Kutta solver, the history thinning procedure, and implementation aspects relevant for performance and reproducibility.

\subsection{Two-dimensional non-uniform grid on the causal triangle}
To discretize the causal domain \( \{(t,t') \mid 0 \le t' \le t \le T_{\rm sim}\} \) in a manner that is both accurate and computationally efficient, we introduce a two-dimensional non-uniform mesh that takes into account the aforementioned expected separation of time scales for the evolution of strongly interacting systems to late times. 

This is achieved by discretizing the ratio $\theta=t'/t\in[0,1]$ 
on a fixed, highly non-equidistant grid of length $L$. As a consequence, slow dynamics on scales increasing with the age of the system are well resolved. On the other hand, the fast microscopic dynamics expected near $\tau=t-t'\sim\tau_\text{micro}$ and, in the common case that the dynamics was started by a sudden quench at $t=0$ near $t'\sim \tau_\text{micro}$, requires the grid to satisfy the constraint
\begin{align}\label{eq:condition}
    \theta_{i+1}-\theta_i\ll \frac{\tau_\text{micro}}{t} \quad \forall i:\min(\theta_i,1-\theta_i)\lesssim\frac{\tau_\text{micro}}{\min{(t',\tau)}}
\end{align}
for all simulated times $t$.

\begin{figure}[t]
\centering
\includegraphics[width=0.75\columnwidth]{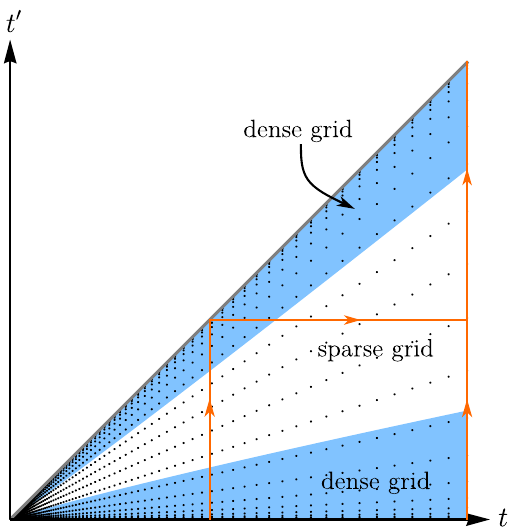}
\caption{The \textsc{Dynamite} time domain is defined by \(t \ge t'\ge 0\). Black dots denote the time grid, which is adaptive in the \(t\) direction and uses a fixed, non-equidistant grid in \(\theta=t'/t\in[0,1]\). The grid is dense near the diagonal and at short times \(t'\ll t\), where the evolution is fast, and sparse in between. Typical integration contours are shown in orange.}
\label{fig:grid}
\end{figure}

With the above choice, the two-time grid points are given by \((t_{n},\,t_{n}\,\theta_{i})\), with the time steps of the first time argument $t=t_n$ controlled by the adaptive ODE solver, see Sec.~\ref{sec:RK}. It thus forms an irregular grid with larger steps where the fields evolve slowly and smaller ones during transients. On the other hand, for each time $t_n$, the grid in \(\theta\)-space is reused, giving a triangular array of nodes that becomes increasingly coarse in absolute \(t'\)-spacing as \(t_{n}\) grows, yet remains dense in relative terms near the critical regions, see Fig.~\ref{fig:grid}.

Although the performance of the method as tested on the spherical $p$-spin model is largely independent of the specific choice of the discretization, the code available at \cite{DYNAMITE2026} implements the following family of discretizations for $i\in 1,\dots,L$
\begin{align}\label{eq:theta}
\begin{split}
    \theta(i)&=\frac{\tan ^{-1}\left(e^{\gamma }\right)-\tan^{-1}\left(e^{-\gamma y(z_{(\alpha,\delta)}(i))}\right)}{\tan^{-1}\left(e^{\gamma }\right)-\tan^{-1}\left(e^{-\gamma }\right)},\\
    y(i)&=\frac{2i-1-L}{L-1},\\
   z_{(\alpha,\delta)}(i)&=\alpha \phi_\delta(i)+(1-\alpha)i,\\
   \phi_\delta(i)&=\frac{L-1}{2}\left(\frac{g_\delta(y(i))}{g_\delta(1)}+1\right)+1,\\
   g_\delta(x)&=\left[\left(x^3+\delta^3\right)^{1/3}-\delta\right],
\end{split}
\end{align}
where $\gamma=-W_{-1}\left(-1/t_\text{max}\right)$ is the Lambert $W$ function with $t_\text{max}$ the largest simulated time.
Here $\alpha\in[0,1]$ and $\delta\ge 0$ are tradeoff parameters that flatten the distribution of grid points for $\tau \gg \tau_\text{micro}$ and $t'\gg \tau_\text{micro}$ without affecting the condition stated in Eq.~\eqref{eq:condition}.

The implementation stores the full history of the correlation and response functions together with their derivatives $\partial_t C(t,t\theta)$ and $\partial_t R(t,t\theta)$. These are directly provided by the ODE solver at no additional cost and are convenient for the interpolation of the history.

\subsection{Efficient interpolation}\label{sec:interpolation}

The accurate evaluation of the memory integrals requires interpolating the two-time functions
$\mathcal{C}(t,\theta)\equiv C(t,t\theta)$ and $\mathcal{R}(t,\theta)\equiv R(t,t\theta)$.
These interpolations must be performed at sufficiently high order to avoid introducing errors that exceed the user-specified tolerance. At the same time, interpolation constitutes the most time-consuming part of the simulation, with performance primarily limited by memory throughput and scaling linearly with the interpolation order. Moreover, higher-order interpolations are more susceptible to oscillations that may amplify numerical errors during the evolution.

In practice, \textsc{Dynamite} interpolates a generic field $\mathcal{A}(t,\theta)$ using a local cubic Hermite interpolation on the adaptive grid in $t$, and provides multiple options for the fixed $\theta$ grid: a barycentric Lagrange interpolation and a Floater--Hormann rational barycentric interpolation~\cite{Floater2007} as well as the same interpolations performed on the equidistant index grid. In the latter case, one uses the fact that the grid is chosen such that $f(i)=f(\theta(i))$ is a smooth function. Since the function $\theta(i)$ is known analytically, index interpolations usually achieve higher accuracy and stability. Hence, the Lagrange index interpolation is chosen as the default. While the latter offers maximal efficiency, the Floater-Hormann interpolation employs a larger stencil to reduce sensitivity to noise. The choice of interpolation scheme and stencil size allows the user to balance accuracy against computational cost; in Sec.~\ref{sec:results} we explicitly verify that, for the default parameters, interpolation errors remain negligible compared to the Runge--Kutta integration error, see Fig.~\ref{fig:interpol_error}.

The memory integrals appearing in Eq.~\eqref{eq:genericDMFE} can be written in the form
\begin{align}
\begin{split}
    \int_0^{t'} \dd s\, \mathcal{A}(t,s) B(t',s)
    &= t \int_0^\theta \dd\phi\,
    \mathcal{A}(t,\phi)\mathcal{B}(\theta t,\phi/\theta)
    \equiv t\, I_1(t,\theta),
    \\
    \int_{t'}^{t} \dd s\, \mathcal{A}(t,s) B(s,t)
    &= t \int_\theta^1 \dd \phi\,
    \mathcal{A}(t,\phi)\mathcal{B}(\phi t,\theta/\phi)
    \equiv t\, I_2(t,\theta).
\end{split}
\end{align}
For their numerical evaluation, it is essential that the discretization of $\phi$ satisfies condition~\eqref{eq:condition} at every step while using a fixed number of sampling points. This is achieved by defining
$\phi^{(1)}_{ij}=\theta_i\theta_j$ and
$\phi^{(2)}_{ij}=\theta_j+(1-\theta_j)\theta_i$,
which automatically satisfy Eq.~\eqref{eq:condition} whenever the $\theta$ grid does.

Each of the integrals must be evaluated at the $L$ points of the $\theta$ grid using $L$ sampling points. Because the highly non-equidistant grid defined in Eq.~\eqref{eq:theta} does not allow for the reuse of sampling locations, this leads to $\mathcal{O}(L^2)$ interpolations per time step, making interpolation the dominant computational cost.

Using an interpolation with stencil size $s$ and precomputed weights $w^{(1,2)}$, we approximate
\begin{align}
    \mathcal{A}(t,\phi^{(1,2)}_{ij})
    \approx
    \sum_{l=1}^L w^{(1,2)}_{ijl}\,\mathcal{A}(t,\theta_l)\equiv \mathcal{A}^{(1,2)}_{ij}(t).
\end{align}
Since the interpolations are local, the corresponding weight tensors are sparse, with only $sL^2$ non-vanishing entries.

For $I_1(t,\theta_j)$ one finds the simplification
\begin{align}
    \mathcal{B}(\theta_j t,\phi^{(1)}_{ij}/\theta_j)
    = \mathcal{B}(\theta_j t,\theta_i)\equiv \mathcal{B}^{(1)}_{ij}(t),
\end{align}
so that only an interpolation in the first argument is required. Because $\partial_t\mathcal{C}$ and $\partial_t\mathcal{R}$ are directly provided by the ODE solver, \textsc{Dynamite} employs a local cubic Hermite interpolation in $t$. For the relatively dense adaptive $t$ grid this is sufficient: defining $\Delta_n=t_{n+1}-t_n$, the integrated interpolation error per step scales as $\sim \Delta_n^5$, and is therefore comparable to the Runge--Kutta error $\epsilon_{\text{RK}}$ of the fourth-order strong stability preserving scheme with $10$ stages, SSPRK(10,4).

In contrast, $I_2(t,\theta_j)$ requires a genuine two-dimensional interpolation,
\begin{align}
    \mathcal{B}(\phi^{(2)}_{ij} t,\theta_j/\phi^{(2)}_{ij})
    =
    \sum_{l=1}^L w^{(3)}_{ijl}\,
    \mathcal{B}\left(\phi^{(2)}_{ij} t,\theta_l\right)\equiv \mathcal{B}^{(2)}_{ij}(t),
\end{align}
which is evaluated using a Lagrange or Floater--Hormann interpolation in $\theta$ with precomputed weights, followed by a cubic Hermite interpolation in $t$.

While the weights for the $\theta$ interpolation can be precomputed, this is not possible for the adaptive $t$ grid. However, at late times, each new time step is small compared to the age of the system, so that successive interpolation points change only weakly. Initializing the interpolation search with the indices from the previous step and exploiting the monotonicity of both grids allows the search to succeed in constant time on CPUs. On GPUs, where this strategy is less efficient to parallelize, a simple binary search initialized with the previous indices is used instead. For all grid sizes considered, the cost of this search remains negligible compared to the interpolation itself.

\subsection{Convolution integrals}

With the interpolation procedures described in Sec.~\ref{sec:interpolation}, all quantities entering the memory kernels are available at the required quadrature nodes. The remaining task is therefore to assemble the convolution integrals themselves, which by design of $\phi^{(1,2)}$ reduces to local weighted sums over the fixed $\theta$ grid.

Concretely, the two contributions introduced in Eq.~\eqref{eq:int} are evaluated as
\begin{align}\label{eq:int}
    I_{1,2}(t,\theta_i)=\sum_{j=1}^L \nu_j\, \mathcal{A}^{(1,2)}_{ij}(t)\, \mathcal{B}^{(1,2)}_{ij}(t)\,.
\end{align}
Here, the weights $\nu_j$, $j=1,\dots,L$, are precomputed once using a global spline interpolation on the $\theta$ grid. As for the interpolations, the order of the quadrature is chosen sufficiently high to ensure that integration errors remain negligible compared to the user-specified tolerance $\epsilon$. By default, \textsc{Dynamite} employs a quintic spline, although other orders may be selected by the user.

Because the quadrature is local in memory and involves only contiguous reads, the evaluation of Eq.~\eqref{eq:int} is significantly cheaper than the interpolations required to obtain $\mathcal{A}^{(1,2)}_{ij}(t)$ and $\mathcal{B}^{(1,2)}_{ij}(t)$. As a result, the overall cost of each time step is dominated by interpolation rather than by convolution assembly.

Both the interpolation stage and the subsequent convolution assembly are embarrassingly parallel over the $\theta$ grid and consist primarily of simple arithmetic combined with indirect memory accesses. As a result, the overall performance is largely memory-bandwidth limited. This computational structure maps naturally onto GPU architectures, where the large degree of data parallelism and substantially higher memory throughput can be exploited effectively. In practice, GPU acceleration leads to order-of-magnitude reductions in wall-clock time per time step for the models considered here; representative benchmarks are presented in Sec.~\ref{sec:results}.

\subsection{Adaptive ODE solver}\label{sec:RK}

The discussion above provides an efficient procedure to evaluate the right-hand side of Eq.~\eqref{eq:genericDMFE} at any given time $t$. The remaining task is to propagate the correlation, and response functions $C(t,t\theta_i)$ and $R(t,t\theta_i)$ forward in $t$. The objective is to achieve this with the minimal number of convolutions and interpolations while maintaining a prescribed accuracy. The stringent precision requirements favor high-order ODE solvers, whereas the goal of minimizing interpolations at late times motivates methods with large stability domains.

To analyze the local stability of the integration, we linearize the discretized DMFE on the fixed $\theta$-grid $\{\theta_i\}_{i=1}^N$. Denoting
\begin{align}
\begin{split}
    \mathbf{C}(t) = \big(C(t,\theta_1 t), \dots, C(t,\theta_N t)\big)^T,\\
    \mathbf{R}(t) = \big(R(t,\theta_1 t), \dots, R(t,\theta_N t)\big)^T,
\end{split}
\end{align}
the Jacobian $J(t)$ is defined via the linearized evolution of infinitesimal perturbations $(\delta \mathbf{C}, \delta \mathbf{R})$:
\begin{align}
    \frac{\mathrm{d}}{\mathrm{d}t} 
    \begin{pmatrix} \delta \mathbf{C}(t) \\ \delta \mathbf{R}(t) \end{pmatrix} 
    = J(t)\,
    \begin{pmatrix} \delta \mathbf{C}(t) \\ \delta \mathbf{R}(t) \end{pmatrix},
\end{align}
where $J(t)$ is obtained by differentiating the discretized right-hand side of Eq.~\eqref{eq:genericDMFE} with respect to $(\mathbf{C},\mathbf{R})$.

For the (mixed) spherical $p$-spin models, the spectrum of $J(t)$ is tightly confined to the negative real axis, with spectral radius bounded by
\begin{align}
    \lVert J\rVert \le 4\sqrt{f''(1)}\,.
\end{align}
Combined with the stability domain of the chosen Runge--Kutta method, this bound determines the maximal stable time step for the integration.

In principle, the desire for large stability domains would suggest the use of (semi-)implicit Runge--Kutta schemes. However, aging implies that the equations of motion are not only stiff but also increasingly ill-conditioned, due to excitations that become gapless at asymptotically late times. As a consequence, even approximate Jacobian inversions become numerically challenging. Since, in addition, the small error tolerances typically restrict the practically usable time steps, \textsc{Dynamite} employs exclusively explicit ODE solvers.

Although Runge--Kutta methods with extended stability domains for stiff differential equations exist \cite{Vaquero2016,Abdulle2002,OSullivan2015,Medovikov1998,Ketcheson2012}, these are of limited utility for the partial integro-differential DMFE arising in glassy dynamics  
because the numerical solver must satisfy a modified strong stability preservation condition that prevents the growth of unphysical oscillations. 
Consequently, schemes with smaller stability domains and fewer, ideally equidistant, stages tend to perform better in practice. This consideration is reflected in \textsc{Dynamite}, which uses the Dormand--Prince method \cite{Dormand1980} at short times and switches to a fourth-order strong-stability-preserving method with ten stages, SSPRK(10,4), at later times \cite{Ketcheson2008,Fekete2022}. For situations that are less constrained by strong stability preservation, a class of second-order stabilized extended Runge--Kutta methods is also available \cite{Vaquero2009}.

The time evolution is initialized with the Dormand--Prince method until the time step grows sufficiently to reach the boundary of its stability domain. At that point, \textsc{Dynamite} temporarily halves the time step and continues with SSPRK(10,4). For the slow evolution characteristic of the $p$-spin models, a simple adaptive strategy is sufficient. Both Runge--Kutta methods provide error estimates for the correlation and response functions, denoted $\delta C$ and $\delta R$. From these, we compute the $1$-norm
\begin{align}
\epsilon_{\text{RK}}=\sum_i \bigl\lVert \delta C(t,\theta_i t)\bigr\rVert_1 + \bigl\lVert \delta R(t,\theta_i t)\bigr\rVert_1,
\end{align}
and require $\epsilon_{\text{RK}}<\epsilon$, where $\epsilon$ is the user-specified tolerance. If $\epsilon_{\text{RK}}>\epsilon$, the next time step is reduced by a factor of $0.9$. If $\epsilon_{\text{RK}}<\epsilon/2$, it is increased by a factor of $1.01$. Finally, if $\epsilon_{\text{RK}}>2\epsilon$, the previous ten time steps are reverted and the time step is halved.

\subsection{Selective history reduction}\label{sec:sparsify}

\begin{figure}[t]
  \centering
  \includegraphics[width=0.8\columnwidth]{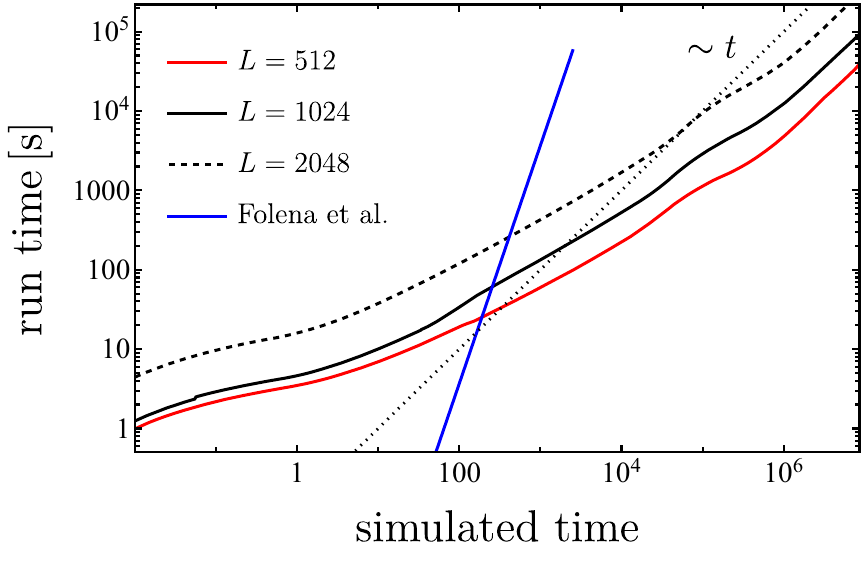}
  \includegraphics[width=0.8\columnwidth]{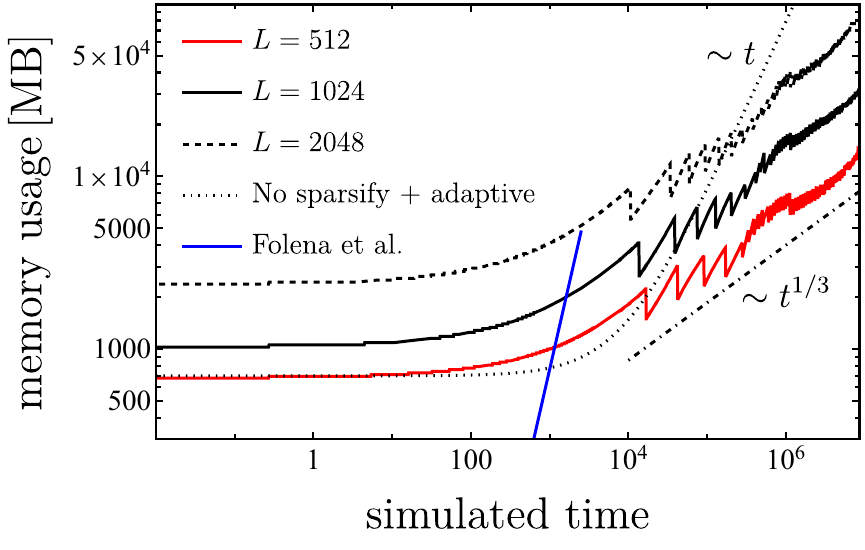}
  \caption{Upper panel: Run time as a function of simulated time for quenches of the mixed spherical $3+4$-spin model with $\lambda=1/2$ using grid lengths $L=512$, $L=1024$, and $L=2048$. For comparison, we show the cubic scaling obtained by using an equidistant grid~\cite{Folena2020} (blue). The dotted line corresponds to a linear scaling of wall-clock time, which is the asymptotic scaling of \textsc{Dynamite}. Lower panel: Total graphics memory used by the algorithm for the same parameters. The dotted line represents the behavior for $L=512$ with constant time step $\Delta t=0.1$ and without sparsification. The dashdotted line approximates the asymptotic behavior with a scaling $\sim t^{1/3}$. All runs were performed with an NVIDIA H100 GPU.}
  \label{fig:performance}
\end{figure}

If the full history of computed time steps were retained, the memory footprint would grow asymptotically linearly with the simulated time. On graphics cards with limited device memory, this would eventually prevent access to late times, see Fig.~\ref{fig:performance}. To avoid this, \textsc{Dynamite} periodically sparsifies the stored history.

Sparsification proceeds locally in time. For each time slice $t_n$, the code evaluates the contribution of this slice to the integrated correlation and response functions over the interval $[t_{n-1},t_{n+1}]$. Concretely, using a cubic Hermite spline, the integrals are computed both with and without the $n$-th slice. Denoting by $\tilde C$ and $\tilde R$ the interpolated functions obtained after removing this slice, we define the local sparsification residual
\begin{align}
\Delta_n(t,\theta_i)
=
\int_{t_{n-1}}^{t_{n+1}} \dd t\,
\Big[(C,R)(t,t\theta_i)-(\tilde C,\tilde R)(t,t\theta_i)\Big],
\end{align}
and set
\begin{align}
\epsilon_{\text{sparse}}
=
\sum_{i=1}^L \bigl\lVert \Delta_n(t,\theta_i) \bigr\rVert_1 .
\end{align}
If $\epsilon_{\text{sparse}}<\epsilon/10$, where $\epsilon$ denotes the user-specified global tolerance, the slice at $t_n$ is discarded and the procedure continues with the $(n+2)$-th slice. Otherwise, the slice is retained, and the algorithm advances to the next candidate. This ensures that the sparsification error remains negligible compared to the Runge--Kutta error $\epsilon_{\text{RK}}$, preventing the occurrence of artifacts in the subsequent evolution following history sparsification.

The overall degree of sparsification can be controlled by the user by specifying the number of sparsification passes performed during the evolution.

\section{Test case and benchmark}
\label{sec:results}

In this section, we provide examples and benchmarks for the use of \textsc{Dynamite}. By comparing with known exact results for the spherical $p$-spin models (with $p=2$ and $p=3$), we demonstrate the high accuracy that \textsc{Dynamite} achieves at unprecedented late times. We benchmark the performance of the GPU implementation and demonstrate various ways in which the simulation results can be used.

\subsection{Performance}
\textsc{Dynamite} uses an explicit adaptive high-order Runge-Kutta method to propagate the equations of motion forward in time. The numerical complexity of each time step is constant. As a result, the CPU time scales sublinearly with the simulated time until the size of the time step reaches the limits of the stability domain of the Runge-Kutta method. From that point on, the computational cost scales linearly with the simulated time. Due to the controlled sparsification of the history, the memory footprint scales sublinearly at all times. In our tests on the mixed spherical $p$-spin model, we find its approximate asymptotic scaling to be $\sim t^{1/3}$. Both of these behaviors are illustrated in Fig.~\ref{fig:performance}. 

\begin{figure}[t]
\centering
\includegraphics[width=0.8\columnwidth]{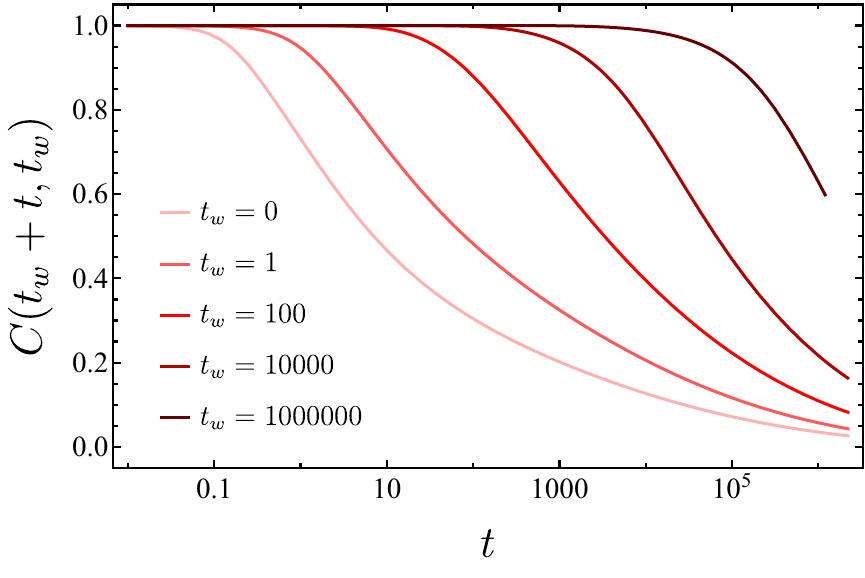}
\caption{The correlation function $C(t_w+t,t_w)$ in the spherical $3+9$-spin model with $\lambda=1/2$ for various waiting times $t_w$ shows the hallmark of aging: As the system grows older (larger $t_w$), the evolution slows down.} 
\label{fig:waiting}
\end{figure} 

\subsection{Accuracy}
\begin{figure}[t]
\centering
\includegraphics[width=0.8\columnwidth]{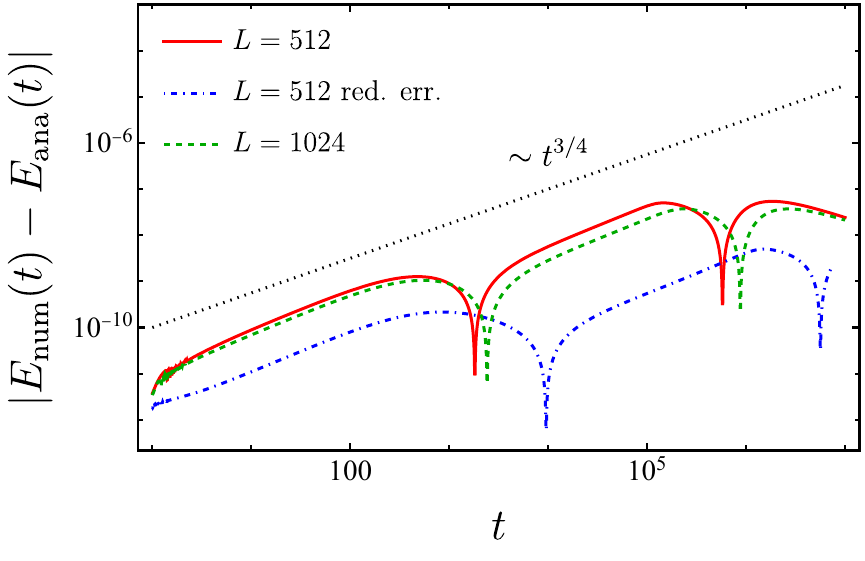}
\caption{The difference between the simulated and the exact energy of the pure 2-spin model following a quench from $T=\infty$ to $T=0$. The blue dash-dotted curve was obtained with a reduced error bound \texttt{-e 1e-12}.} 
\label{fig:2-accuracy}
\end{figure}

The dynamics of aging systems slows down as the system ages, as Fig.~\ref{fig:waiting} clearly shows. Consequently, the Jacobian of glassy systems is ill-conditioned, with the smallest eigenvalues at most proportional to the inverse age of the system. Prevention of the uncontrolled growth of the associated excitations places stringent bounds on the accuracy of the numerical solver. Roughly speaking, stability to time $t$ requires relative errors $\epsilon = \mathit{o}(t^{-1})$ for all integrals. Simultaneously, the memory integrals get longer, causing the effect of stochastic noise to grow $\sim t$. Consequently, reaching late times requires consistently small errors of the interpolation, integration, Runge-Kutta, and sparsification routines. We demonstrate \textsc{Dynamite}'s control of the total error using the exactly solvable spherical $p=2$ spin model in Fig.~\ref{fig:2-accuracy}. 

\begin{figure}[t]
\centering
\includegraphics[width=0.8\columnwidth]{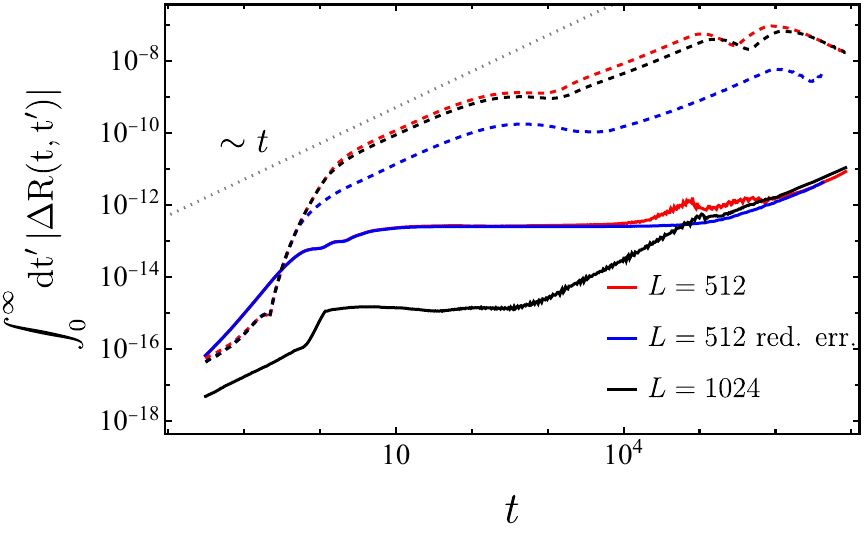}
\caption{Estimated interpolation error of the default 9th-order Lagrange interpolation for the response function of the pure $p=2$ spin model (solid lines), compared with the total deviation between numerical and analytical solutions (dashed lines).}
\label{fig:interpol_error}
\end{figure}

Fig.~\ref{fig:interpol_error} shows the accumulated deviation
$\int_0^t \dd t'\, |R_{\text{ana}}(t,t')-R_{\text{num}}(t,t')|$
together with the corresponding estimate of the interpolation error for the 9th-order index Lagrange interpolation scheme. In all cases the interpolation error remains negligible throughout the evolution, confirming that interpolation does not constitute the dominant source of inaccuracy in the parameter ranges considered here.

\begin{figure}[t]
\centering
\includegraphics[width=0.8\columnwidth]{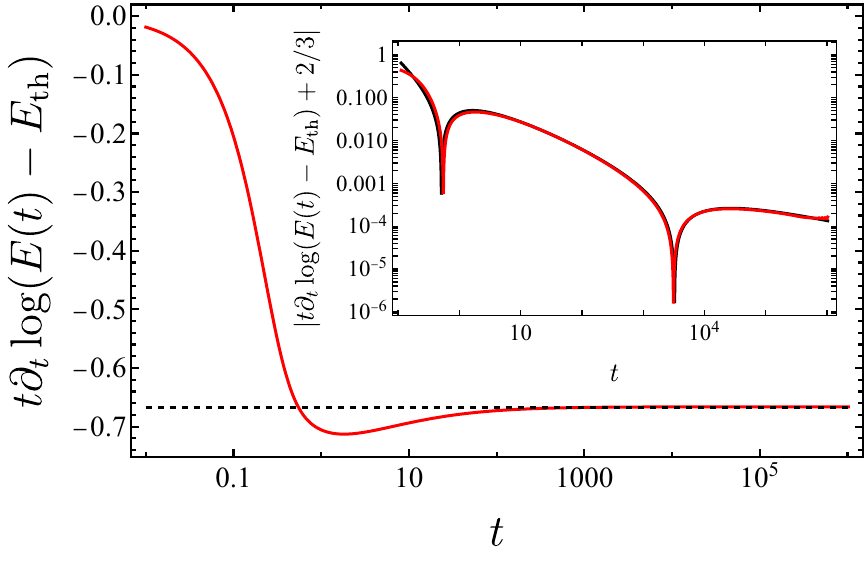}
\caption{Convergence of the energy exponent in the pure $p=3$ model. The asymptotic behavior $E(t)-E_\text{th}\sim t^{-2/3}$ is recovered with very high accuracy using the command line argument \texttt{-e 1e-12}. The inset shows the rate of convergence of the exponent (red) together with a fit of the form $\sum_{i=1}^3 \alpha_i t^{-i/3}$ (black curve). This suggests that asymptotically $E(t)-E_\text{th}$ is a series in powers of $t^{-1/3}$.}
\label{fig:exponent}
\end{figure}

We further demonstrate the accuracy achievable with \textsc{Dynamite} by investigating in the spherical $p=3$ spin model the exponent of the algebraic convergence of the energy $E(t)$ to its asymptotic value  \cite{Folena2020}
\begin{equation}
    E_\text{th}=\frac{f(1)-f'(1)}{\sqrt{f''(1)}}-\frac{f(1)\sqrt{f''(1)}}{f'(1)}=-\sqrt{\frac{2(p-1)}{p}}\;,
    \label{eq:Eth}
\end{equation}
(where the first expression is generic and the second one holds only for pure models) following a quench from $T_0=1/\beta_0=\infty$ to $T=0$.
We define $\delta E(t)=E(t)-E_\text{th}$ and we assume asymptotically $\delta E(t) \sim t^{-\alpha}$.
In Fig.~\ref{fig:exponent} we plot
\begin{align*}
t \partial_t \log \delta E(t)=\frac{t\, \delta E'(t)}{\delta E(t)} \to -\alpha\;.
\end{align*}
We notice the convergence to the value $\alpha=2/3$, which is supported by the analytical argument reported in the following paragraph.
Moreover, a fit in powers of $t^{-1/3}$ interpolates the data in the whole range (see inset in Fig.~\ref{fig:exponent}).

The analytical argument predicting the exponent $\alpha=2/3$ for the decay of the energy in pure spherical $p$-spin models goes as follows.
It is well known that, in these pure models, the spectrum of the Hessian computed at a stationary point of energy $E=E_\text{th}+\epsilon$ follows a semi-circular law with a lower band edge in $\lambda_\text{min} \propto -\epsilon$. Thus, the fraction of negative eigenvalues scales like
\[
\int_{\lambda_\text{min}}^0 \sqrt{\lambda-\lambda_\text{min}}\; \dd\lambda \propto \epsilon^{3/2}\;.
\]
A large times the relaxation process is slow because the system configuration is, with high probability, very close to a stationary point, where the gradient is null. In such a situation, the stochastic evolution is close to a random walk (because of the almost flat space) and the time it takes the system to proceed further in the energy relaxation is inversely proportional to the fraction of negative directions, $\tau(\epsilon) \sim \epsilon^{-3/2}$. Inverting such a relation one gets $\epsilon \sim t^{-2/3}$, that is $\alpha=2/3$.

\begin{figure}[t]
  \centering
  \includegraphics[width=0.8\linewidth]{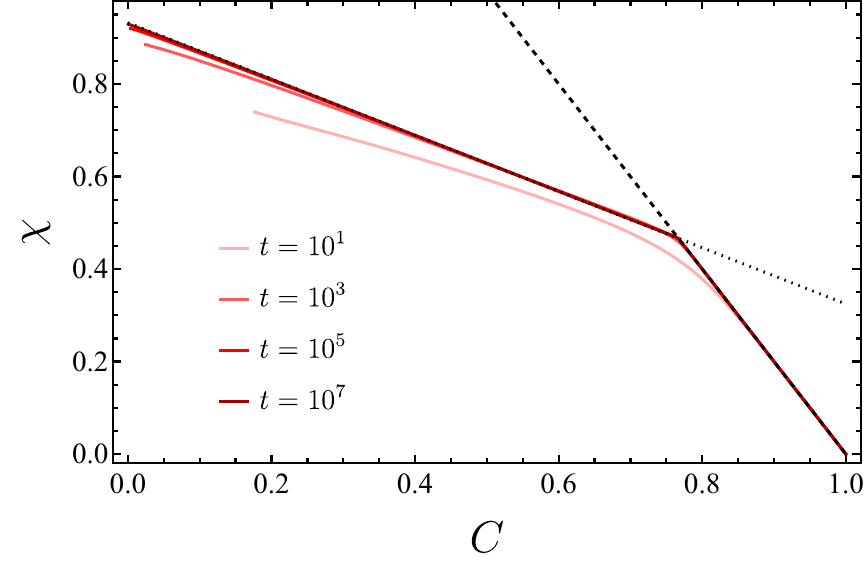}
  \includegraphics[width=0.8\linewidth]{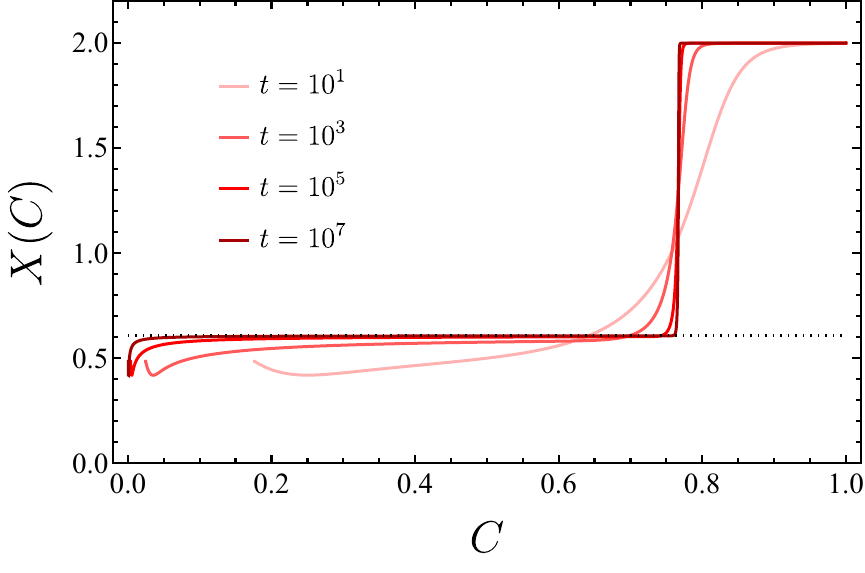}
  \caption{Upper panel: Integrated response $\chi(t,s)=\int_s^t \dd\tau R(t,\tau)$ versus correlation $C(t,s)$ for various $t$ values, measured in the pure $p=3$ spin model following a quench from $T_0=\infty$ to $T=1/2$. For $C>q_\text{\tiny EA} \approx 0.766911$, the slope corresponds to the inverse bath temperature $\beta=1/T=2$ (dashed line). For $C<q_\text{\tiny EA}$ the slope converges to $x_\text{th} \approx 0.607865$ as predicted by the exact solution. Lower panel: The numerical effective inverse temperature as a function of the correlation.}
  \label{fig:RvsC}
\end{figure}

\subsection{Aging dynamics}
The hallmark of aging dynamics is the emergence of an effective inverse temperature $X[C]$ defined via \cite{Cugliandolo1994}
\begin{align}
R(t,t')=X[C(t,t')]\partial_{t'}C(t,t')\,.
\end{align}
In the case of the pure spherical $p$-spin model, in the limit of large times $X$ approaches the constant value $x_\text{th}$ in the aging regime (i.e., for $C(t,t')<q_\text{\tiny EA}$, being $q_\text{\tiny EA}$, the Edwards-Anderson order parameter). For $T=0$ its value is
\[
x_\text{th}=\frac{\sqrt{f''(1)}}{f'(1)}-\frac{1}{\sqrt{f''(1)}}=\sqrt{\frac{2}{p(p-1)}}(p-2),
\]
while for generic temperatures it can be obtained from the analytic solution to the model \cite{Crisanti1992}.
The output of our integration scheme reproduces perfectly the expected behavior (see Fig.~\ref{fig:RvsC}).

\begin{figure}[t]
\centering
\includegraphics[width=0.8\columnwidth]{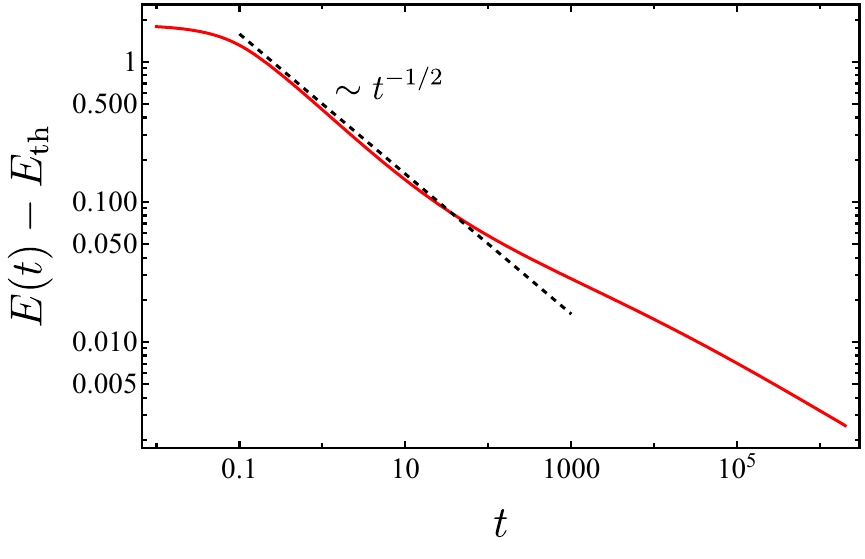}
\caption{For mixed spherical spin models, the energy typically relaxes much more slowly than in pure models. Here we show the quench from $T_0=\infty$ to $T=0$ of the $3+9$ model with $\lambda=1/2$. Mixed models commonly show an initial fast relaxation (here $\sim t^{-1/2}$) followed by a slower relaxation. Discussion of the asymptotic dynamics thus requires access to much later times than for the pure models.}
\label{fig:mixedEnergyConvergence}
\end{figure}

While the dynamics of the pure spherical $p$-spin models are well understood, the same cannot be said for mixed models. For example, it is known \cite{Folena2023} that, for some mixtures with a large value of $\lambda$, the threshold energy in Eq.~\eqref{eq:Eth} drops below the lowest energy achievable by an algorithm in algebraic time \cite{Alaoui2020}. Consequently, $X[C]$ cannot be a constant, but its analytic form is yet unknown. Even in cases where $X[C]$ converges to a constant, the evolution is much slower than in pure models, as can be seen in Fig.~\eqref{fig:mixedEnergyConvergence}, and estimating the decay exponent requires a more detailed analysis that we leave for a future work~\cite{Lang2026}.

\section{Discussion and conclusions}
\label{sec:discussion}

In this work, we have presented \textsc{Dynamite}, a powerful numerical framework for the integration of dynamical mean-field equations involving two-time correlation and response functions. The method combines a non-uniform discretization of the causal triangle with high-order local interpolations, explicit adaptive Runge--Kutta time stepping, and controlled thinning of the stored history. Together, these ingredients enable long-time integration of non-stationary DFME while keeping both computational cost and memory usage manageable.

The central algorithmic strategy is to exploit the multiscale structure of two-time dynamics by representing the fields on a fixed grid in the relative coordinate $\theta=t'/t$ and an adaptive grid in the physical time $t$. This allows interpolation weights and quadrature coefficients associated with $\theta$ to be precomputed once, so that each time step reduces to local interpolations and weighted sums over the stored history. Combined with adaptive time stepping and periodic removal of redundant past slices, this substantially reduces the number of kernel evaluations required during the evolution.

As anticipated in Sec.~\ref{sec:thframework}, this approach is particularly advantageous in situations where
\begin{enumerate}[label=(\roman*)]
\item the dynamics is slow, requiring access to very long times while retaining the full two-time history, and/or
\item the evaluation of the memory kernels $K$ and $D$ is itself computationally expensive.
\end{enumerate}

Beyond the spherical $p$-spin models used here as benchmarks, the numerical strategy implemented in \textsc{Dynamite} applies to a broad class of problems governed by causal two-time integro-differential equations of the form~\eqref{eq:genericDMFE}. 
Importantly, the algorithm is not tied to the specific structure of glassy dynamics, but rather to the generic features of dynamical mean-field equations: causality on a triangular domain, history-dependent kernels, and the coexistence of fast microscopic and slow collective time scales. These characteristics arise in many classical and quantum Dynamical Mean-Field Theory (DMFT) formulations.

As a concrete example, closely related equations appear in nonequilibrium DMFT for correlated lattice systems, where two-time Green's functions evolve under self-consistent memory kernels. In such settings, including recent studies of driven or quenched Hubbard-type models, the present framework could be employed with minimal modifications by replacing the model-specific kernels while retaining the same numerical infrastructure. This highlights that \textsc{Dynamite} should be viewed as a general-purpose integrator for long-time DMFT dynamics rather than a solver specialized to a particular model.

The code is released with CPU and GPU backends, checkpointing, and provenance tracking to facilitate reproducible simulations and adaptation to new models.

From a numerical perspective, the accuracy is maintained through a combination of Runge--Kutta error control, high-order interpolation, and conservative thinning criteria. Nevertheless, some limitations should be noted. While errors are controlled locally at each step, the interpolation does not enforce physical constraints such as $\partial_{t'}C(t,t')>0$ or $R(t,t')>0$. In certain parameter regimes, this may lead to unphysical oscillations that can be amplified during late-time evolution. Such effects could likely be mitigated by interpolation schemes that are less sensitive to overshoots~\cite{Akima1970}, which we leave for future work.

Looking ahead, several technical extensions of \textsc{Dynamite} appear natural. In its current form, GPU performance is often limited by available device memory. This restriction could be alleviated by employing higher-order interpolation schemes on the adaptive $t$ grid, which would allow a further reduction of stored history points. In addition, subtracting analytically known contributions to the late-time behavior near the diagonal may enable the use of larger time steps, thereby extending the accessible simulation times even further. Finally, while the results presented here were obtained for classical dynamics, the same algorithmic structure applies directly to quenched quantum systems formulated in real time, which are out of reach for Langevin-based approaches.

In summary, \textsc{Dynamite} provides a practical tool for integrating two-time dynamical mean-field equations in regimes that are difficult to access with uniform discretizations. By reducing the effective cost of history-dependent convolutions while maintaining controlled accuracy, it enables long-time simulations in systems with slow dynamics and/or expensive kernels, and offers a flexible starting point for further methodological developments in this direction.

\section*{Acknowledgments}
We thank Sebastian Diehl for fruitful discussions and insightful feedback on the manuscript, and Subir Sachdev for valuable collaboration on earlier work that contributed to this publication.
The work of J.L.\ was supported by the Deutsche Forschungsgemeinschaft (DFG, German Research Foundation) under Germany’s Excellence Strategy Cluster of Excellence Matter and Light for Quantum Computing (ML4Q) EXC 2004/1 390534769, and by the DFG Collaborative Research Center (CRC) 183 Project No. 277101999, as well as the DFG CRC 1238 Project No. 277146847.
The work of F.R.T.\ was supported by the “National Centre for HPC, Big Data and Quantum Computing”, Project CN\_00000013, CUP B83C22002940006, NRRP Mission 4 Component 2 Investment 1.4,  Funded by the European Union - NextGenerationEU.

\appendix

\section{Implementation details and usability}\label{sec:implementation}

\textsc{Dynamite} \cite{DYNAMITE2026} provides a complete CPU implementation written in C++ together with an optional CUDA backend selected at runtime. Several compute-intensive components, including interpolation, convolution, sparsification, and selected time-stepping routines, are available on both CPU and GPU. The CPU path remains available for reference and validation runs, with hot loops parallelized using \texttt{OpenMP} when available. On GPUs, kernels are organized to favor coalesced memory access and data reuse, with scalar model parameters stored in constant memory and small kernels distributed across multiple streams to mitigate launch overhead.

Long simulations are checkpointed either in HDF5 format (when available) or in a compact binary fallback format. Checkpointing may be performed synchronously or asynchronously, with at most one write in flight at any time. In addition to full-state checkpoints, the code exports compressed snapshots intended for post-processing workflows that do not require the complete history at full resolution. 

To facilitate reproducibility, each run produces a human-readable parameter record that includes the full command line, version information (Git hash and build state), compiler and CUDA versions, basic system information, and the grid-genera\-tion parameters associated with the selected precomputed grid. Core metadata is also embedded directly into HDF5 checkpoints when used.

For improved robustness of the response channel, \textsc{Dynamite} optionally supports logarithmic interpolation of the response history. When enabled, interpolation is performed on $\log R$ (and the corresponding scaled derivative), with automatic fallback to linear interpolation whenever stencil values become non-positive. This option is available on both CPU and GPU paths.

The project includes a complete user-facing documentation
(available at \cite{DYNAMITE2026})
consisting of usage references, conceptual pages, and tutorials.
To facilitate analysis and figure generation, lightweight Python scripts are provided to process the compressed snapshot files. Together, these features are intended to lower the barrier to applying the solver to new models and platforms, while maintaining reproducibility and providing consistent numerical behavior across architectures.

\section{Numerical parameter selection}

The discretization parameters $L$, interpolation order $m$, spline degree $s$ (if used), and the Runge--Kutta tolerance should be chosen based on standard convergence tests with respect to grid resolution and solver accuracy. In practice, these parameters are adjusted until relevant observables and two-time functions remain stable within the desired precision.

The maximal accessible time scale is typically limited by the accuracy of the interpolated memory integrals. Small interpolation errors accumulate over time and are amplified when high interpolation orders are used. Once amplified, these errors force the adaptive Runge--Kutta solver to reduce the time step progressively until the minimal step size is reached, at which point the integration terminates. Such instabilities are readily identified by high-frequency noise on the $\theta$ grid of both correlation and response functions.

The onset of this behavior can be delayed by lowering the Runge--Kutta error tolerance, increasing the grid length $L$, and reducing the interpolation order. In practice, index interpolation (the default method) is considerably more stable in this respect than interpolation performed directly on the non-equidistant $\theta$ grid.

The implementation is thread-safe and parallelized using OpenMP for interpolation and convolution tiles. Optional CUDA support enables device acceleration. Build options allow portable compilation across heterogeneous systems.

\bibliographystyle{elsarticle-num}
\bibliography{bibliography}
\end{document}